\documentclass[a4paper]{amsart}
\usepackage{geometry}
\usepackage{graphicx}
\usepackage[font=scriptsize]{caption}
\usepackage{tikz}
\usetikzlibrary{trees,positioning}

\usepackage[T1]{fontenc}

\RequirePackage{amsmath}
\RequirePackage{bm}
\RequirePackage{amssymb}
\RequirePackage{upref}
\RequirePackage{amsthm}
\RequirePackage{enumerate}
\RequirePackage{pb-diagram}
\RequirePackage{amsfonts}
\RequirePackage[mathscr]{eucal}
\RequirePackage{verbatim}
\RequirePackage{xr}
\RequirePackage{graphicx}
\usepackage{calc}
\usepackage{xspace}
\RequirePackage{color}
\RequirePackage{ifthen}
\RequirePackage{fancybox}
\RequirePackage{mathrsfs}

\newcommand{\cf}{cf.\@\xspace}

\newcommand{\al}{\alpha}
\newcommand{\bet}{\beta}
\newcommand{\ga}{\gamma}
\newcommand{\de}{\delta }

\newcommand{\f}{\varphi}

\newcommand{\ka}{\kappa}
\newcommand{\lam}{\lambda}

\newcommand{\s}{\sigma}

\newcommand{\vsi}{\varsigma}

\newcommand{\C}{\varGamma}
\newcommand{\D}{\varDelta}
\newcommand{\F}{\varPhi}
\newcommand{\Lam}{\varLambda}

\newcommand{\so}{{\mc S_0}}
\newcommand{\socc}{{\mc S_0}}

\newcommand{\const}{\tup{const}}

\newcommand{\msp[1]}[1]{\mspace{#1mu}}

\newcommand{\R}[1][n+1]{{\protect\mathbb R}^{#1}}

\newcommand{\Hh}[1][n+1]{{\protect\mathbb H}^{#1}}
\newcommand{\Ss}[1][n+1]{{\protect\mathbb S}^{#1}}

\newcommand{\N}{{\protect\mathbb N}}

\newcommand{\eR}{\stackrel{\lower1ex \hbox{\rule{6.5pt}{0.5pt}}}{\msp[3]\R[]}}
\newcommand{\eN}{\stackrel{\lower1ex \hbox{\rule{6.5pt}{0.5pt}}}{\msp[1]\N}}
\newcommand{\eO}{\stackrel{\lower1ex \hbox{\rule{6pt}{0.5pt}}}{\msc O}}

\DeclareMathOperator{\tr}{tr}

\newcommand\ra{\rightarrow}

\newcommand\pa{\partial}
\newcommand\pde[2]{\frac {\partial#1}{\partial#2}}
\newcommand{\un}{\infty}
\newcommand{\A}{\forall}

\newcommand{\uu}{\cup}
\newcommand{\ii}{\cap}
\newcommand{\uuu}{\bigcup}

\newcommand{\uud}{ \stackrel{\lower 1ex \hbox {.}}{\uu}}
\newcommand{\uuud}[1]{ \stackrel{\lower 1ex \hbox {.}}{\uuu_{#1}}}
\newcommand\su{\subset}

\newcommand{\sminus}[1][28]{\raise 0.#1ex\hbox{$\scriptstyle\setminus$}}

\newcommand{\wed}{\wedge}

\newcommand{\abs}[1]{\lvert#1\rvert}

\newcommand{\spd}[2]{\protect\langle #1,#2\protect\rangle}

\newcommand{\tit}{\textit}

\newcommand{\tup}{\textup}% text upright

\newcommand{\mc}{\protect\mathcal}
\newcommand{\msc}{\protect\mathscr}

\newcommand{\tlam}{\tilde\lam}

\newcommand{\bt}{\begin{thm}}
\newcommand{\bl}{\begin{lem}}
\newcommand{\bc}{\begin{cor}}
\newcommand{\bd}{\begin{definition}}
\newcommand{\bpp}{\begin{prop}}
\newcommand{\br}{\begin{rem}}
\newcommand{\bn}{\begin{note}}
\newcommand{\be}{\begin{ex}}
\newcommand{\bes}{\begin{exs}}
\newcommand{\bb}{\begin{example}}
\newcommand{\bbs}{\begin{examples}}
\newcommand{\ba}{\begin{axiom}}
\newcommand{\bas}{\begin{assumption}}

\newcommand{\et}{\end{thm}}
\newcommand{\el}{\end{lem}}
\newcommand{\ec}{\end{cor}}
\newcommand{\ed}{\end{definition}}
\newcommand{\epp}{\end{prop}}
\newcommand{\er}{\end{rem}}
\newcommand{\en}{\end{note}}
\newcommand{\ee}{\end{ex}}
\newcommand{\ees}{\end{exs}}
\newcommand{\eb}{\end{example}}
\newcommand{\ebs}{\end{examples}}
\newcommand{\ea}{\end{axiom}}
\newcommand{\eas}{\end{assumption}}

\newcommand{\bp}{\begin{proof}}
\newcommand{\ep}{\end{proof}}
\newcommand{\eps}{\renewcommand{\qed}{}\end{proof}}

\newcommand{\bal}{\begin{align}}

\newcommand{\bi}[1][1.]{\begin{enumerate}[\upshape #1]}
\newcommand{\bia}[1][(1)]{\begin{enumerate}[\upshape #1]}
\newcommand{\bin}[1][1]{\begin{enumerate}[\upshape\bfseries #1]}
\newcommand{\bir}[1][(i)]{\begin{enumerate}[\upshape #1]}
\newcommand{\bic}[1][(i)]{\begin{enumerate}[\upshape\hspace{2\cma}#1]}
\newcommand{\bis}[2][1.]{\begin{enumerate}[\upshape\hspace{#2\parindent}#1]}
\newcommand{\ei}{\end{enumerate}}

\newcommand\ndots{\raise 0.47ex \hbox {,}\hskip0.06em\cdots %
     \raise 0.47ex \hbox {,}\hskip0.06em} 

\newcommand{\q}{\quad}
\newcommand{\qq}{\qquad}

\newcommand\nd{\noindent}

\newskip\Csmallskipamount                                                
\Csmallskipamount=\smallskipamount
\newskip\Cmedskipamount
\Cmedskipamount=\medskipamount
\newskip\Cbigskipamount
\Cbigskipamount=\bigskipamount

\newcommand\cvs{\vspace\Csmallskipamount}   
\newcommand\cvm{\vspace\Cmedskipamount}

\newskip\csa
\csa=\smallskipamount

\newskip\cma
\cma=\medskipamount

\newskip\cba
\cba=\bigskipamount

\newdimen\spt
\newcommand\citem{\cvs\advance\itemno by
1{(\romannumeral\the\itemno})\hskip3pt}
\newcommand{\bitem}{\cvm\nd\advance\itemno by
1{\bf\the\itemno}\hspace{\cma}}

\newcount\itemno
\newcommand{\las}[1]{\label{S:#1}}

\newcommand{\lae}[1]{\label{E:#1}}
\newcommand{\lat}[1]{\label{T:#1}}
\newcommand{\lal}[1]{\label{L:#1}}

\newcommand{\lac}[1]{\label{C:#1}}

\newcommand{\lar}[1]{\label{R:#1}}

\newcommand{\rs}[1]{Section~\ref{S:#1}}

\newcommand{\rt}[1]{Theorem~\ref{T:#1}}
\newcommand{\rl}[1]{Lemma~\ref{L:#1}}

\newcommand{\rr}[1]{Remark~\ref{R:#1}}

\newcommand{\re}[1]{\eqref{E:#1}}
\newcommand{\frc}[1]{Corollary~\ref{C:#1} on page~\tup{\pageref{C:#1}}}
\newcommand{\frt}[1]{Theorem~\ref{T:#1} on page~\tup{\pageref{T:#1}}}
\newcommand{\frl}[1]{Lemma~\ref{L:#1} on page~\tup{\pageref{L:#1}}}

\newcommand{\fre}[1]{\eqref{E:#1} on page~\tup{\pageref{E:#1}}}

\newskip\thmskip
\thmskip=\parindent

\newskip\hsk
\newenvironment{hinw}{\labelsep=0pt\begin{list}{}{\labelsep=0pt\itemindent=0pt\labelwidth=0pt\leftmargin=\parindent\rightmargin=0pt\partopsep=\cba}%
\item\it\nopagebreak\nopagebreak}%
{\end{list}}

\newcommand\bh{\begin{hinw}}
\newcommand{\eh}{\end{hinw}}

\newtheoremstyle{normal}% name
  {\cba}%      Space above, empty = `usual value'
  {\cba}%      Space below
  {}% Body font
  {\thmskip}%Indent amount (empty = no indent, \parindent = para indent)
  {\bfseries}% Thm head font
  {.}%        Punctuation after thm head
  {\hsk}%     Space after thm head: " " = normal interword space;
  {}% Thm head spec

\newtheoremstyle{abschnitt}% name
  {\cba}%      Space above, empty = `usual value'
  {\cba}%      Space below
  {}% Body font
  {\thmskip}% Indent amount (empty = no indent, \parindent = para indent)
  {\bfseries}% Thm head font
  {.}%        Punctuation after thm head
  {\hsk}%     Space after thm head: " " = normal interword space;
  {}% Thm head spec

\newtheoremstyle{italic}% name
  {\cba}%      Space above, empty = `usual value'
  {\cba}%      Space below
  {\itshape}% Body font
  {\thmskip}%  Indent amount (empty = no indent, \parindent = para indent)
  {\bfseries}% Thm head font
  {.}%        Punctuation after thm head
  {\hsk}%     Space after thm head: " " = normal interword space;
  {}% Thm head spec

\newtheoremstyle{aufgaben}% name
  {\cba}%      Space above, empty = `usual value'
  {\cba}%      Space below
  {}% Body font
  {}%         Indent amount (empty = no indent, \parindent = para indent)
  {\normalsize\bfseries}% Thm head font
  {.}%        Punctuation after thm head
  {\hsk}%     Space after thm head: " " = normal interword space;
  {}% Thm head spec

\newtheoremstyle{break}% name
  {\cba}%      Space above, empty = `usual value'
  {\cba}%      Space below
  {\itshape}% Body font
  {}%         Indent amount (empty = no indent, \parindent = para indent)
  {\bfseries}% Thm head font
  {.}%        Punctuation after thm head
  {\newline}% Space after thm head: \newline = linebreak
  {}%         Thm head spec

\theoremstyle{italic}
\newtheorem{thm}[subsection]{Theorem}
\newtheorem{lem}[subsection]{Lemma}
\newtheorem{prop}[subsection]{Proposition}
\newtheorem{cor}[subsection]{Corollary}

\theoremstyle{normal}
\newtheorem{rem}[subsection]{Remark}
\newtheorem{definition}[subsection]{Definition}
\newtheorem{example}[subsection]{Example}
\newtheorem{examples}[subsection]{Examples}
\newtheorem{ex}[subsection]{Exercise}
\newtheorem{note}[subsection]{}
\newtheorem{axiom}[subsection]{Axiom}
\newtheorem{assumption}[subsection]{Assumption}

\theoremstyle{aufgaben}
\newtheorem{exs}[subsection]{Exercises}

\numberwithin{equation}{section}
\numberwithin{figure}{section}

\newenvironment{textequation}[1][0.8]
{\begin{equation}
\begin{aligned}
\begin{minipage}{#1\linewidth}}
{\end{minipage}
\end{aligned}
\end{equation}
\ignorespacesafterend}

\newcommand{\btext}{\begin{textequation}}
\newcommand{\etext}{\end{textequation}}

\def\hinweis{\@startsection{subsection}{2}%
 \z@{0.7\linespacing\@plus 0.5\linespacing}{0.7\linespacing}%
{\normalfont\itshape\indent}}

\newcounter{hours}\newcounter{minutes}
\newcommand{\printtime}{%
\setcounter{hours}{\time/60}%
\setcounter{minutes}{\time-\value{hours}*60}%
\ifthenelse{\value{minutes}<10}{\thehours :0\theminutes}{\thehours:\theminutes}}

\usepackage[german,english]{babel}
\usepackage{graphicx}
\RequirePackage{amsmath}
\RequirePackage{bm}
\RequirePackage{amssymb}
\RequirePackage{upref}
\RequirePackage{amsthm}
\RequirePackage{enumerate}
\RequirePackage{pb-diagram}
\RequirePackage{amsfonts}
\RequirePackage[mathscr]{eucal}
\RequirePackage{verbatim}
\RequirePackage{xr}
\RequirePackage{graphicx}
\RequirePackage{mathrsfs}
\usepackage{calc}
\usepackage{xspace}
\usepackage{dsfont}

\makeatletter
\RequirePackage{color}
\newcommand{\ann}[1]{\renewcommand{\@makefnmark}{\mbox{$^{\color{red}{\@thefnmark}}$}}%
\footnote {#1}}
\makeatother

\RequirePackage{upref}
\RequirePackage{amsthm}
\RequirePackage{enumerate}%\begin{enumerate}[(i)]
\usepackage[mathscr]{eucal}
\usepackage{xr-hyper}

\usepackage{calc}

\newlength{\oddsidemarginlength}
\newlength{\topmarginlength}

\newcounter{numberoflines}
\newcounter{tempcc}
\oddsidemargin=\oddsidemarginlength
\evensidemargin=\oddsidemargin
\topmargin=\topmarginlength
\usepackage[colorlinks=true,linkcolor=blue,citecolor=blue,urlcolor=blue]{hyperref}  

\begin{document}

\flushbottom

%\larger[1]
%\frontmatter

\title[Quantizing the exterior region of a black hole]{Quantizing the exterior region of a Schwarzschild-AdS black hole leads to a resolution of the information paradox on a quantum level}

% author one information
\author{Claus Gerhardt}
\address{Ruprecht-Karls-Universit\"at, Institut f\"ur Angewandte Mathematik,
Im Neuenheimer Feld 205, 69120 Heidelberg, Germany}
%\curraddr{}
\email{\href{mailto:gerhardt@math.uni-heidelberg.de}{gerhardt@math.uni-heidelberg.de}}
\urladdr{\href{http://www.math.uni-heidelberg.de/studinfo/gerhardt/}{http://www.math.uni-heidelberg.de/studinfo/gerhardt/}}
%\thanks{This work was supported by the DFG}

% author two information
%\author{}
%\address{}
%\curraddr{}
%\email{
%\thanks{}
%
%\subjclass[2000]{35J60, 53C21, 53C44, 53C50, 58J05}
%\keywords{Lorentzian manifold, mass, cosmological spacetime, general relativity, inverse mean curvature flow, ARW spacetimes}


\date{\today}
%
% at present the "communicated by" line appears only in ERA and PROC
%\commby{}

%\dedicatory{}

\begin{abstract} 
We quantize the exterior region of a Schwarzschild-AdS black hole using our model of quantum gravity. The resulting hyperbolic equation is solved by products of temporal eigenfunctions $w_i$, the eigenvalues of which all have multiplicity one, and spatial eigendistributions $v_{ij}$ having the same eigenvalues but with multiplicities $1\le m_i$, where the $m_i$ could in principle be arbitrarily large. Regarding only the exterior region, there was no guidance how to determine the values of the $m_i$. However, considering also the quantization of the interior region, where the same question did not arise since the $m_i$ could be chosen by maximizing the value, it seemed logical to choose the same values, too, in the exterior case. Since the eigenvalues in the interior are the same because the temporal Hamiltonian is the same in both cases, this choice defined a unitary equivalence between the respective Hilbert spaces and the respective Hamiltonians. Hence, there is no information paradox on a quantum level.
\end{abstract}

\maketitle

\tableofcontents

\setcounter{section}{0}
\section{Introduction}

In \cite[Chapter 7]{cg:qgravity-book2} we applied our model of quantum gravity to the interior of a Schwarzschild-AdS black hole $N$ of dimension $n+1$, $n\ge 3$. For the quantization we worked in a fiber bundle $E$ with base space $\so$. The fiber elements are the Riemannian metrics $g_{ij}$ in $\so$, which could be expressed, in a suitable coordinate system, in the form 
\begin{equation}
g_{ij}(t,x)=t^\frac4n \s_{ij}(x)\qq\A\, x\in \so,
\end{equation}
where $0<t<\un$. The fibers $F(x)$ are globally hyperbolic manifolds with respect to the DeWitt metric and $t$ is a time function independent of $x$. The metrics $\s_{ij}(x)$ belong to a subbundle $M$
\begin{equation}
M=\{t=1\}\su E.
\end{equation}

Picking a Cauchy hypersurface $\so$ with induced metric $\s_{ij}$ in $N$ its quantum development would be governed by the hyperbolic equation, \cf equation \cite[(4.2.53), p.~115]{cg:qgravity-book2}, namely
\begin{equation}\lae{6.1.1}
\begin{aligned}
&\frac n{16(n-1)}t^{-(m+k)}\frac\pa {\pa t}(t^{(m+k)}\dot u)\\
&\q -t^{-2}\D_Mu+\frac12 t^{-2}\D_{\R[k]}u-(n-1)t^{2-\frac4n}\D_\s u\\
&\q-(\frac n2-1)t^{2-\frac4n}R_\s u +(n-2)t^2\Lam u=0,
\end{aligned}
\end{equation}
where $u$ depends on $(x,t,\s_{ij},\theta)$, with $x\in\so$, $t\in \R[]_+$, $\s_{ij}\in M$ and $\theta\in \R[k]$. The equation is \tit{evaluated} at $(x,t,\s_{ij},\bar\theta)$, where $(x,t)$ are variables but $\s_{ij}$ is the fixed metric of the Cauchy hypersurface $\so$ after evaluation and
\begin{equation}
\bar\theta^a(x)=1\qq\A\,x\in \so\; \tup{and}\; 1\le a\le k.
\end{equation}
Let us recall that $\s_{ij}$ may be considered to be an element of $M$, in view of \cite[Remark 3.2.2, p.~75]{cg:qgravity-book2}. The Laplacian $\D_\s$ is the Laplacian with respect to $\s_{ij}$ after fixing, $R$ is the scalar curvature of the metric, $0<t$ is the time coordinate defined by the derivation process of the equation and $\Lam<0$ a cosmological constant. Finally, $\D_M$ may be identified with the Laplacian in the symmetric space  
\begin{equation}
SL(n,\R[])/SO(n),
\end{equation}
in view of \cite[Lemma 3.2.1, p.~74]{cg:qgravity-book2}. We proved that the hypersurface $\so$ is a product space 
\begin{equation}
\so=\R[]\times M_0,
\end{equation}
where $M_0$ is a compact Riemannian manifold and that $\s$ is a product metric
\begin{equation}\lae{2.6.1.5.1}
\s=\de\otimes \bar\s,
\end{equation}
where $\de$ is the "metric" in $\R[]$ and $\bar\s$ the metric in $M_0$. Indeed, in the Schwarzschild case $M_0$ will be a space form with curvature
\begin{equation}
\tilde\ka\in\, \{-1,0,1\}.
\end{equation}
The metric in \re{2.6.1.5.1} is free of any coordinate singularity. Since $\so$ is a Cauchy hypersurface in the black hole region
\begin{equation}
0<r<r_0,
\end{equation}
i.e., a slice 
\begin{equation}
\so=\{r=\const\},
\end{equation}
where $r_0$ is the radius of the event horizon, $\bar\s$ depends on $r$ and it converges smoothly to a Riemannian metric if $r$ tends to $r_0$. 

In this paper we want to quantize the \tit{exterior} region of a Schwarzschild-AdS black hole. The formal quantization is the same as for the interior region. However, the Cauchy hypersurfaces $\so$ are then the slices
\begin{equation}
\so=\{t=\const\}
\end{equation}
where $t$ is the time coordinate in the exterior region. The Cauchy hypersurfaces for different values of $t$ are all isometric, with the induced metric being independent of $t$ and can be expressed as
\begin{equation}\lae{1.12}
ds^2=d\tau^2+r(\tau)^2\tilde\s_{ij}dx^idx^j,
\end{equation}
where the coordinate $\tau$ ranges over the interval $[0,\un)$, and $\tau=0$ corresponds to the event horizon. The Cauchy hypersurface $\so$ can then be identified with the product
\begin{equation}
\so=(0,\un)\times M_0
\end{equation}
equipped with the metric given above. The scalar curvature of the metric is equal to $2\Lam$, \cf \frl{2.1}. Hence the solution $u$ of \re{6.1.1} can then be considered to be a product of eigenfunctions or eigendistributions of the various differential operators, i.e., we look for solutions   
\begin{equation}
u=w\hat v\vsi v
\end{equation}
where $w=w(t)$ is a temporal solution, $\hat v=\hat v(\s_{ij})$ an eigenfunction of $-\D_M$, $\vsi=\vsi(\theta^a)$ should be an eigenfunction of $-\D_{\R[k]}$ and $v=v(\tau,x)$ an eigenfunction of $-\D_\s$, the Laplacian of $\so$. For $\hat v$ we choose one of the elementary gravitons which are eigenfunctions of $-\D_M$ satisfying
\begin{equation}\lae{2.8.3.27}
-\D_M {\hat v}=(\abs\lam^2+\abs\rho^2)\hat v
\end{equation}
and
\begin{equation}\lae{2.8.3.28}
\hat v(\s (x))=1\qq\A\, x\in \socc,
\end{equation}
where $\s(x)$ is the induced metric of the Cauchy hypersurface $\so$ and
\begin{equation}
\abs{\rho}^2=\frac{(n-1)^2n}{12},
\end{equation}
\cf \cite[Lemma 3.2.3, p.~76 \& equ. (2.2.40), p.~49]{cg:qgravity-book2}. 
The function $\vsi$ is defined by $\vsi=1$ such that the scalar field produces no contribution to the solution apart from increasing the dimension. 

If we consider solutions $u$ of \re{6.1.1} with these choices of eigenfunctions then the equation \re{6.1.1} reduces to
\begin{equation}\lae{1.18}
(H_0w)v-(H_1v)w=0,
\end{equation}
where $H_0$ is the temporal Hamiltonian 
\begin{equation}\lae{1.19}
\begin{aligned}
H_0w&=t^{-(2-\frac4n)}\bigg(-\frac n{16(n-1)} t^{-(m+k)}\frac\pa{\pa t}\big (t^{(m+k)} \pde wt\big )\\
&\msp[100] -t^{-2}(\abs\lam^2+\rho^2)w -(n-2)t^2\Lam w\bigg)
\end{aligned}
\end{equation}
with $t$ ranging in $0<t<\un$, and $H_1$ is the spatial Hamiltonian
\begin{equation}
\begin{aligned}
H_1v&=-(n-1)\D_\s v -(\frac n2-1) R_\s v\\
&=-(n-1)\D_\s v -(n-2) \Lam v,
\end{aligned}
\end{equation}
since $R_\s=2\Lam$.
\br\lar{1.1}
In \cite[Theorem 3.5.5]{cg:qgravity-book2} we proved, under reasonable assumptions, that $H_0$ is self-adjoint in an appropriate Hilbert space with a pure point spectrum and eigenfunctions $w_i$ satisfying
\begin{equation}
H_0w_i=\lam_iw_i\qq\A\, i\in \N
\end{equation}
such that
\begin{equation}
0<\lam_0<\lam_1<\lam_2<\cdots,
\end{equation}
\begin{equation}
\lim_{i\ra \un}\lam_i=\un,
\end{equation}
\begin{equation}
\lim_{t\ra 0}\abs{w_i(t)}=\un
\end{equation}
and
\begin{equation}
\lim_{t\ra\un}\abs{w_i(t)}=0.
\end{equation}
Moreover, for any $\bet>0$ the operator
\begin{equation}
e^{-\bet H_0}
\end{equation}
is of trace class. Hence, we considered the singularity at $t=0$ as a big bang on a quantum level. In \cite{cg:qgravity5} we could show that the eigenfunctions $\tilde u_i$ of a unitarily equivalent operator, which are defined by
\begin{equation}
\tilde u_i =t^\frac{m+k}2 w_i,
\end{equation}
can be smoothly extended past the singularity by even or odd mirroring. This extension also applies to the corresponding equations such that the extended eigenfunctions can be looked at as classical solutions of these equations valid even in $t=0$.
\er

The main part of the spatial Hamiltonian $H_1$ is the Laplacian of the Cauchy hypersurface $\so$. We shall prove in the next section that its eigenfunctions or eigendistributions are products of the eigenfunctions of the compact space forms $M_0$ and of the eigenfunctions of a self-adjoint differential operator on the interval $(0,\un)$ which we shall call $A_0$. The operator is bounded from below, has a continuous spectrum and no square integrable eigenfunctions with positive eigenvalues. But it has distributional eigenfunctions which are smooth, bounded and hence tempered distributions.

In \rs{3} we define a sequence of eigenfunctions $v_{ij}$ of $H_1$ satisfying
\begin{equation}\las{1.28}
H_1 v_{ij}=\lam_i v_{ij},\qq 1\le j\le m_i,
\end{equation}
where $\lam_i$ are the eigenvalues of $H_0$ which all have multiplicity one, but as eigenvalues of $H_1$ they have multiplicity $m_i$. The $m_i$ are arbitrary predetermined natural numbers. 

In \rs{4} we apply quantum statistics to our configuration by choosing $H_1$ as the underlying self-adjoint operator, since the eigenvalues of $H_0$ have multiplicity one, i.e., the spatial Hamiltonian is the right choice in order to define the partition function, the von Neumann entropy and the average energy. When we quantized the interior region of a Schwarzschild-AdS black hole we had a similar situation. However, then we solved this problem by maximizing the multiplicity $m_i$. For each $i\in\N$ the maximal value of $m_i$ was bounded from above by the geometric settings of the Cauchy hypersurfaces 
\begin{equation}
\so=\{r=\const\}
\end{equation}
and the $m_i(r)$ increased when $r$ tended to $r_0$, the radius of the event horizon. In the exterior region it makes no sense to maximize multiplicities since the $m_i$ can be prescribed arbitrarily; they are unbounded from above. Thus, the only logical and physically meaningful choice is to use the same value for each $m_i$ as in the interior region. This postulation then defines a unitary operator between the respective Hilbert spaces and makes the interior and exterior spatial Hamiltonians unitarily equivalent, i.e., their quantum statistics are identical. Hence, the black hole information paradox does not exist on a quantum level.

We shall also prove:
\bt
The spatial eigenfunctions $v_{ij}=v_{ij}(\tau,x)$ can be looked at as being gravitational waves emanating from the event horizon and vanishing exponentially fast at infinity satisfying 
\begin{equation}
v_{ij}(0,x)=0
\end{equation}
and
\begin{equation}
\lim_{\tau\ra \un} \abs{v_{ij}(\tau,\cdot)}_{m,M_0}=0\qq\A\,m\in \N,
\end{equation}
where we use the norm in $C^m(M_0)$. Furthermore, $v'_{ij}$ also vanishes exponentially fast at infinity such that
\begin{equation}
\sup_{x\in M_0}\int_0^\un\abs{v_{ij}(\tau,x)}^2+\abs{v'_{ij}(\tau,x)}^2<\un.
\end{equation}
\et

Similar results are also valid for the quantization of a Kerr-AdS black hole, where we already quantized the interior black hole region in \cite[Chapter 8]{cg:qgravity-book2}, compare also an older paper \cite{cg:qbh2}. The arguments are similar to those we present in this paper though some of the proofs are technically more difficult. We intend to write up this work in the next months.

\section{Spacelike slices in the exterior region}\las{2}

The metric in the exterior region of the Kerr-AdS black hole can be expressed in the form 
\begin{equation}\lae{2.1}
d\bar s^2=- hdt^2+h^{-1} dr^2  + r^2\bar\s_{ij}dx^idx^j,
\end{equation}
where $(\bar\s_{ij})$ is the metric of an $(n-1)$-dimensional space form $M_0$ and $h(r)$ is defined by
\begin{equation}\lae{2.2}
 h=-mr^{-(n-2)}-\tfrac2{n(n-1)}\Lam r^2 +\tilde\ka,
\end{equation}
where $m>0$ is the mass of the black hole (or a constant multiple of it), $\Lam<0$ a cosmological constant, and $\tilde\ka\in\{-1,0,1\}$ is the
curvature of $M_0=M_0^{n-1}$, $n\ge 3$. We also stipulate that $M_0$ is compact in the cases $\tilde\ka\not=1$. If $\tilde\ka =1$ we shall assume
\begin{equation}
M_0=\Ss[n-1]
\end{equation}
which is of course the important case. By assuming $M_0$ to be compact we can use eigenfunctions instead of eigendistributions when we consider spatial eigenvalue problems, but we could also consider $M_0=\R[n-1]$ or $M_0=\Hh[n-1]$ with the corresponding eigendistributions for the respective Laplacians.

The radial variable $r$ ranges between
\begin{equation}
r_0<r<\un,
\end{equation}
where $r_0$ is the radius of the \tit{unique} {event horizon}, and the time variable $t$ is an element of an open interval $I=(t_1,t_2)$.

The exterior region of the black hole is a globally hyperbolic $(n+1)$-dimen\-sional spacetime and the hypersurfaces
\begin{equation}
S_t=\{t_1<t=\const<t_2\}
\end{equation}
are Cauchy hypersurfaces with induced metric
\begin{equation}\lae{2.6}
ds^2= h^{-1}dr^2 +r^2\bar\s_{ij}dx^idx^j,
\end{equation}
i.e., the Cauchy hypersurfaces $S_t$ are all isometric and their second fundamental form $h_{a b}$ vanishes. Hence, we infer
\bl\lal{2.1}
The scalar curvature $R$ of the hypersurfaces $S_t$ is constant
\begin{equation}
R=2\Lambda.
\end{equation}
\el
\bp
Since the black hole satisfies the Einstein equations
\begin{equation}
G_{\al\bet}+\Lam \tilde g_{\al\bet}=0
\end{equation}
and $h_{ab}=0$ we deduce from the Gau{\ss} equation
\begin{equation}
R=-\{H^2-\abs A^2\}+2G_{\al\bet}\nu^\al\nu^\bet=2G_{\al\bet}\nu^\al\nu^\bet,
\end{equation}
\cf \cite[equ. (1.1.43), p.\,5]{cg:cp}, completing the proof.
\ep

The exterior region is defined by
\begin{equation}
h>0,
\end{equation}
hence the coordinate transformation 
\begin{equation}\lae{2.11}
\tau=\int_{r_0}^rh^{-\frac12}
\end{equation}
is well defined such that
\begin{equation}\lae{2.12}
\pde r\tau=h^\frac12
\end{equation}
and the metric in \re{2.6} can be expressed in the coordinates $(\xi^a)$=$(\tau,x^i)$ as
\begin{equation}\lae{2.13}
ds^2=d\tau^2+r(\tau)^2\tilde\s_{ij}dx^idx^j.
\end{equation}
Let us denote the metric by $(g_{ab})$ and let $\D$ be the corresponding Laplacian, $\tilde\D$ the Laplacian with respect to $(\tilde\s_{ij})$, $g$ and $\tilde\s$ the respective determinants, then
\begin{equation}
\sqrt g=r^{n-1}\sqrt{\tilde\s}
\end{equation}
and for a function $v=v(\tau, x^i)$ of class $C^2$ we deduce 
\begin{equation}\lae{2.15}
\D v=\frac1{r^{n-1}}\frac{\pa}{\pa\tau}\big(r^{n-1}v'\big)+r^{-2}\tilde\D v,
\end{equation}
where the prime indicates partial differentiation with respect to $\tau$.

We want to solve the eigenvalue equation 
\begin{equation}\lae{2.16}
\D v=-\mu v
\end{equation}
by separation of variables. Thus, let $\f$ be an eigenfunction of $\tilde \D$
\begin{equation}\lae{2.17}
-\tilde \D \f=\lam \f
\end{equation} 
then we define $v$ by
\begin{equation}\lae{2.18}
v=\psi(\tau)\f(x)
\end{equation}
such that $v$ is an eigenfunction if and only if $\psi$ satisfies
\begin{equation}\lae{2.19}
\frac1{r^{n-1}}\frac{\pa}{\pa\tau}\big(r^{n-1}\psi'\big)-\lam r^{-2}\psi=-\mu \psi,
\end{equation}
or equivalently,
\begin{equation}\lae{2.20}
A_{n-1}\psi=-\frac1{r^{n-1}}\frac{\pa}{\pa\tau}\big(r^{n-1}\psi'\big)+\lam r^{-2}\psi=\mu \psi.
\end{equation}
The operator $A_{n-1}$ can be looked at as a densely defined symmetric operator in a suitable Hilbert space.
\bd
Let $I$ be the interval $I=(0,\un)$ and for real valued functions $\psi_1,\psi_2\in C^\un_c(I)$ define the scalar product
\begin{equation}
\spd{\psi_1}{\psi_2}_{n-1}=\int_I\psi_1\psi_2 r^{n-1}d\tau,
\end{equation}
where $r=r(\tau)$ is the function implicitly defined in \re{2.11}. Furthermore, denote the completion of this scalar product space by $L^2(I,n-1)$. Sometimes we shall also write $L^2(I,r^{n-1}d\tau)$ to provide the explicit definition.

The operator $A_{n-1}$ defined in \re{2.20} is a densely defined symmetric operator with corresponding bilinear form
\begin{equation}
\spd{\psi_1}{\psi_2}_2=\spd{A_{n-1}\psi_1}{\psi_2}_{n-1}\qq\A\, \psi_1,\psi_2\in C^\un_c(I).
\end{equation}
The right-hand side of this equation can be written as an integral 
\begin{equation}
\int_I\{\psi_1'\psi_2' r^{n-1}+\lam r^{n-3}\psi_1\psi_2\}d\tau=\spd{\psi_1'}{\psi_2'}_{n-1}+\spd{r^{-2}\psi_1}{\psi_2}_{n-1}.
\end{equation}
\ed
\bl
The operator $A_{n-1}$ is bounded from below
\begin{equation}
A_{n-1}\ge -c_0
\end{equation}
if $c_0=c_0(r_0,\lam)$ is large enough.
\el 
\bp
Obvious, since 
\begin{equation}
0<r_0\le r(\tau)\qq\A\,\tau\in I.
\end{equation}
\ep
Hence, $A_{n-1}$ is essentially self-adjoint and has a unique self-adjoint extension which is known as the Friedrichs extension, \cf \cite[Chapter V.4, p. 110]{maurin:book}.

\br
The previous definitions and results can immediately be generalized to complex functions by defining the scalar products appropriately.
\er
Next, let us define a unitarily equivalent operator the eigenfunctions, or better, eigendistributions of which can be better analyzed. First, we write the function $\psi$ in \re{2.19} as a product
\begin{equation}\lae{2.26}
\psi=r^{-\frac{n-1}2} u,
\end{equation}
where $u=u(\tau)$ is a real function of class $C^2$. Differentiating and simplifying the resulting equation
yields
\begin{equation}
\begin{aligned}
&u''-r^{-2} \big\{\lam -\frac{n-1}2(1+\frac{n-1}2)\abs{r'}^2+\frac{n-1}2r r''\big\}u+\mu u=\\
&u''-r^{-2}\big\{\lam -\frac{n-1}2(1+\frac{n-1}2) h+\frac{n-1}4r\pde hr\big\}u+\mu u=0,
\end{aligned}
\end{equation}
where we used \re{2.12}. From the definition of $h$ in \re{2.2} we deduce
\begin{equation}
\begin{aligned}
\frac12 r\pde hr&=\frac12 m(n-2)r^{-(n-2)}-\frac 2{n(n-1)}\Lam r^2\\
&= h+\frac n2 m r^{-(n-2)}-\tilde\ka,
\end{aligned}
\end{equation}
hence $u$ should satisfy the ODE
\begin{equation}\lae{2.29}
u''-r^{-2}\big\{\lam -\frac {(n-1)^2}4 h+\frac{n(n-1)}4 mr^{-(n-2)}-\frac{(n-1)}2\tilde\ka\big\}u=-\mu u,
\end{equation}
or equivalently,
\begin{equation}\lae{2.30}
-u''+r^{-2}\big\{\lam -\frac {(n-1)^2}4 h+\frac{n(n-1)}4 mr^{-(n-2)}-\frac{(n-1)}2\tilde\ka\big\}u=\mu u.
\end{equation}
The left-hand side of this equation defines a linear differential operator $A_0$
\begin{equation}\lae{2.31}
A_0u=-u''+r^{-2}\big\{\lam -\frac {(n-1)^2}4 h+\frac{n(n-1)}4 mr^{-(n-2)}-\frac{(n-1)}2\tilde\ka\big\}u,
\end{equation}
which is unitarily, or orthogonally, equivalent to $A_{n-1}$ defined in \re{2.20}.
\bl\lal{2.5}
Let $L^2(I)=L^2(I,d\tau)$ be the usual Hilbert space of square integrable real valued functions $u$ defined in $I$, with the standard scalar product
\begin{equation}
\spd{u_1}{u_2}=\int_I u_1 u_2d\tau,
\end{equation}
then the map
\begin{equation}
\begin{aligned}
\F:L^2(I,n-1)&\ra L^2(I)\\
\psi&\ra r^{\frac{(n-1)}2}\psi
\end{aligned}
\end{equation}
is orthogonal and the operators $A_{n-1}$ and $A_0$, the domains of which are the test functions $C^\un_c(I)$, are orthogonally equivalent
\begin{equation}
A_{n-1}=\F^{-1}\circ A_0\circ \F.
\end{equation}
\el
\bp Follows immediately from the arguments after the ansatz \re{2.26}.
\ep

$A_0$ is densely defined, symmetric, bounded from below and hence essentially self-adjoint. We shall prove later that the eigenvalue equation
\begin{equation}\lae{2.35}
A_0u=\mu u
\end{equation}
has no square integrable solutions and that $A_0$ has only a continuous spectrum such that we can only find eigendistributions. At the moment we shall treat the eigenvalue equation simply as an ODE which can be uniquely solved, for any $\mu\in \R[]$, by a function $u\in C^2[0,\un)$ with initial values
\begin{equation}\lae{2.36}
u(0)=0\q\wed\q u'(0)=1.
\end{equation}
In order to analyze the solution of the ODE we observe that the zero order term of $A_0$ can be expressed in the form
\begin{equation}\lae{2.37}
\begin{aligned}
r^{-2}\big\{\lam -\frac {(n-1)^2}4 h+&\frac{n(n-1)}4 mr^{-(n-2)}-\frac{(n-1)}2\tilde\ka\big\}\\
&\equiv q-\frac{n-1}{2n}\abs\Lam,
\end{aligned}
\end{equation}
where
\begin{equation}\lae{2.38}
\abs q\le c_0 r^{-2}\qq\A\, t\in [0,\un)
\end{equation}
with a uniform constant $c_0=c_0(m,r_0,n)$. Therefore, we can rewrite the eigenvalue equation \re{2.35} as
\begin{equation}\lae{2.39}
-u''+q u=\big(\mu +\frac{n-1}{2n}\abs\Lam\big) u\equiv k^2 u                                                          
\end{equation}
assuming
\begin{equation}
k>0,
\end{equation}
or equivalently,
\begin{equation}\lae{2.41}
\mu>-\frac{n-1}{2n}\abs\Lam.
\end{equation}
Moreover, combining \re{2.2} and \re{2.12} we deduce that there are positive constants $\ga_0$ and $r_1\ge r_0$ such that
\begin{equation}
\pde{\log r}\tau=r^{-1} h^\frac12\ge \ga_0\abs\Lam ^\frac12\equiv c_1\qq\A\, r\ge r_1,
\end{equation}
where $\ga_0$ is independent of $\abs\Lam$ for any $\tilde \ka$, but $r_1$ only if $\tilde\ka=1$; otherwise, $r_1$ depends on $\abs\Lam$. We conclude further
\begin{equation}
c_1(\tau-\tau_1)\le \log r-\log r_1,
\end{equation}
where $r_1=r(\tau_1)$, from which we infer 
\begin{equation}\lae{2.44}
r^{-1}\le r_1^{-1}e^{c_1\tau_1}e^{-c_1 \tau}=c_2 e^{-c_1\tau}\qq\A\, \tau\ge \tau_1,
\end{equation}
and finally
\begin{equation}\lae{2.45}
r^{-1}\le r_0^{-1}e^{c_1\tau_1}e^{-c_1 \tau}=c_3 e^{-c_1\tau}\qq\A\, \tau\ge 0.
\end{equation}
Applying now the estimate \re{2.38} we deduce
\begin{equation}\lae{2.46}
\abs q\le c e^{-\ga \tau}\qq\A\,\tau\ge0,
\end{equation}
$c,\ga$ are positive constants, i.e., the potential $q$ vanishes exponentially fast near infinity. Hence we can apply known results for the spectrum and eigenfunctions of the operator in \re{2.39}. It is well-known that the eigenvalue equation $\re{2.39}$ has no square integrable solutions if $k>0$, \cf \cite[Theorem 4.1, p. 14]{simon-spectrum}. The assumptions for this result are actually much weaker for the condition
\begin{equation}
\limsup _{\tau\ra\un}\tau\abs q=0
\end{equation}
would be sufficient. Since we also want to prove that the solutions of equation \re{2.29} satisfying the initial conditions \re{2.36} are bounded and, thus, tempered distributions which can be looked at as eigendistributions of the self-adjoint operator, which requires the use of modified Pr\"ufer coordinates, it is rather simple to add a few arguments to show that there are no square integrable solutions.
\bt\lat{2.6}
Let $u\in C^2([0,\un)$ be a non-trivial solution of the equation
\begin{equation}\lae{2.48}
-u''+q u=k^2 u
\end{equation}
with $k>0$, where $q$ satisfies the estimate \re{2.46}, then
\begin{equation}\lae{2.49}
\abs u^2+k^{-2}\abs {u'}^2\le c_0\qq\A\, \tau\ge 0,
\end{equation}
where $c_0$ depends on $k, c_0, \ga$ and the initial values of $u$ in $\tau=0$. Moreover, $u$ is not square integrable.
\et
\bp
We use modified Pr\"ufer coordinates $(R,\theta)$ as in \cite[Section 2, equs. (1.5a), (1.5b), p.\,3]{simon-spectrum} by defining
\begin{equation}\lae{2.50}
\begin{aligned}
u'&=k R(\tau)\cos(\theta(\tau))\\
u&=R(\tau)\sin(\theta(\tau)).
\end{aligned}
\end{equation} 
We immediately deduce
\begin{equation}
\abs u^2+k^{-2}\abs {u'}^2=R^2.
\end{equation}
Furthermore, by differentiating both equations and by applying simple algebraic manipulations we infer
\begin{equation}\lae{2.52}
R'=\frac q{2k} R \sin(2\theta)
\end{equation}
and
\begin{equation}\lae{2.53}
\theta'=k-\frac qk \sin^2\theta.
\end{equation}
In order to prove the estimate \re{2.49} we conclude from \re{2.52} and \re{2.46}
\begin{equation}
\begin{aligned}
\log R-\log R(0)&=\frac1{2k}\int_0^\tau q\sin(2\theta)\\
&\le \frac1{2k} c\int_0^\tau e^{-\ga \tau}=\frac c{2k\ga} (1-e^{-\ga\tau})\le \frac c{2k\ga},
\end{aligned}
\end{equation}
which implies
\begin{equation}
R(\tau)\le R(0) e^\frac c{2k\ga}\qq\A\,\tau\ge 0.
\end{equation}
Using the same reasoning we also deduce
\begin{equation}
\begin{aligned}
\log R-\log R(0)&=\frac1{2k}\int_0^\tau q\sin(2\theta)\\
&\ge -\frac1{2k} c\int_0^\tau e^{-\ga \tau}\ge -\frac c{2k\ga},
\end{aligned}
\end{equation}
implying
\begin{equation}
R(\tau)\ge R(0)e^{-\frac c{2k\ga}}
\end{equation}
and hence
\begin{equation}
\int_0^\un R^2=\int_0^\un(\abs u^2+k^{-2}\abs {u'}^2)=\un.
\end{equation}
On the other hand, if $u$ would be square integrable then $u'$ would also be square integrable, \cf \cite[Lemma 4.2, p.\,14]{simon-spectrum}, completing the proof of the theorem.
\ep
\br
Let $A$ be the operator defined by the left-hand side of the equation \re{2.48} with domain
\begin{equation}
D(A)=C^\un_c(I),
\end{equation}
where $I=(0,\un)$. Assuming $q\ge 0$, which in our case would require that the eigenvalue $\lam$ in \re{2.37} would be sufficiently large, then the quadratic form defined by
\begin{equation}
\spd{Au}{u}=\int_0^\un (\abs {u'}^2+q\abs u^2)d\tau,\qq u\in D(A),
\end{equation}
would be non-negative and actually positive definite, since
\begin{equation}
\spd{Au}u=0
\end{equation}
implies $u\equiv \const$
 and hence $u= 0$ because $u(0)=0$. It is well-known that the self-adjoint extension of $A$, which we denote by $\tilde A$, is defined in
\begin{equation}
D(\tilde A)=H^{1,2}_0(I)\ii H^{2,2}(I),
\end{equation}
where the function spaces on the right-hand side are the usual Sobolev spaces. Any eigenvector $u$ of $\tilde A$ satisfying
\begin{equation}
\tilde A u=0
\end{equation}
would also belong to $C^2([0,\un))$ satisfying $u(0)=0$ and the eigenvalue equation \re{2.48} with $k=0$. Hence $u$ would vanish identically because
\begin{equation}
\spd {\tilde Au}u=\int_0^\un \tilde Au u \,d\tau=0.
\end{equation}
On the other hand, any square integrable function $u\in C^2([0,\un))$ satisfying $u(0)=0$ which solves the equation \re{2.48} with $k=0$ is an element of $D(\tilde A)$, \cf \cite[Lemma 4.2, p.\,14]{simon-spectrum}. Thus, we conclude $u=0$.
\er
As a corollary of \rt{2.6} we obtain
\bc\lac{2.8}
The eigenvalue problem in \re{2.35}
\begin{equation}
A_0u=\mu u
\end{equation}
with initial values $u(0)=0$ and $u'(0)=1$ and eigenvalue $\mu$ satisfying \re{2.41} has no square integrable solutions. As an ODE it has a unique solution $u\in C^2([0,\un))$ which satisfies the estimate \re{2.49}, where
\begin{equation}\lae{2.66}
k^2=(\mu+\frac{n-1}{2n}\abs\Lam),
\end{equation}
in view of \re{2.39}. Hence, the ODE's solution is a tempered distribution and can be looked at as an eigendistribution of the unique self-adjoint extension $\tilde A_0$.
\ec

Combining the results of the corollary above, \rl{2.5} and the equations \re{2.15}--\re{2.20}, then we conclude
\bt\lat{2.9}
The eigenvalue equation \re{2.16}  
\begin{equation}
-\D v=\mu v
\end{equation}
can be solved by defining  
\begin{equation}
v=r^{-\frac{n-1}2} u\f,
\end{equation}
where $\f=\f(x)$ is an eigenfunction of equation \re{2.17} and $u$ a tempered eigendistribution of equation \re{2.35}, provided $\mu$ satisfies \re{2.41}.
\et
\newpage
\nd
Below is a Mathematica generated plot where the eigenfunction solves a structurally equivalent equation and where also an equivalent damping  is used.
\vskip 1cm
\nd
\begin{figure}[h]
\includegraphics[width=0.9\textwidth]{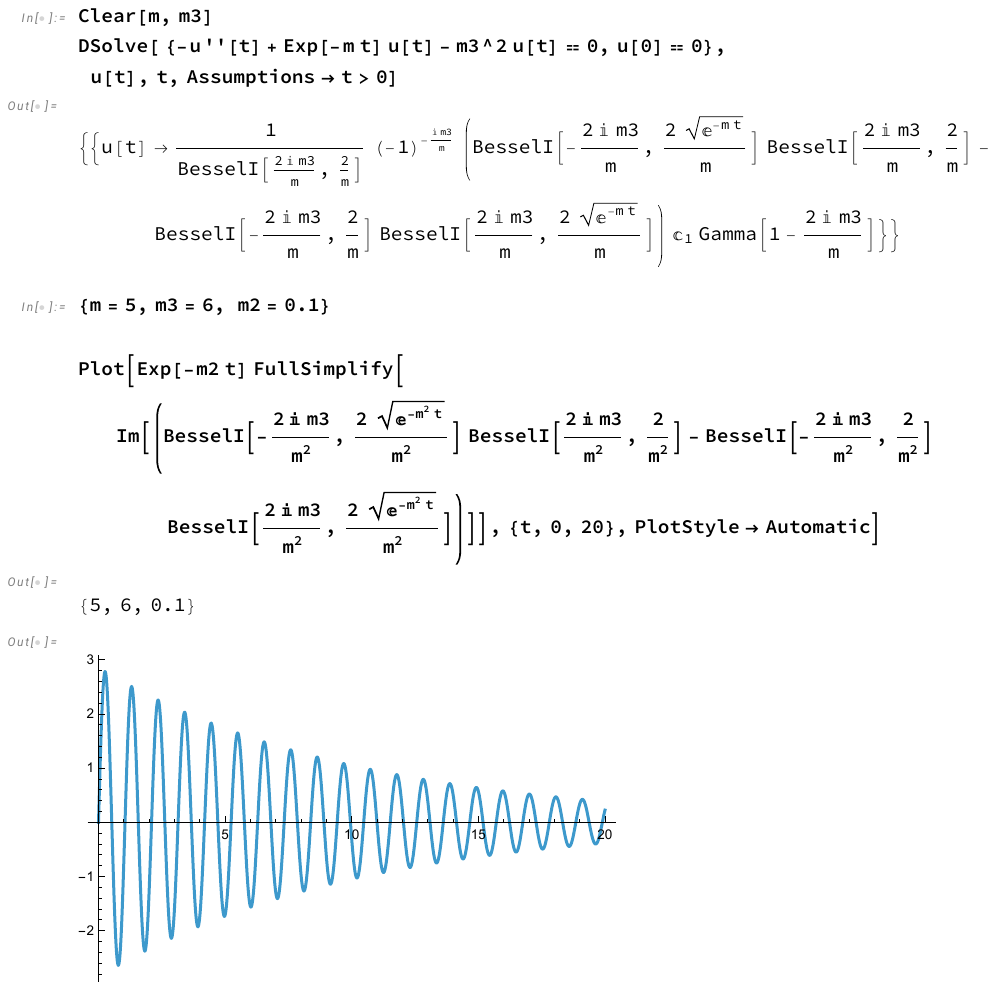}
\end{figure}
%\mbox{}

\section{Solving the hyperbolic equation}\las{3} 
Let us recall from the introduction that the final equation which has to be satisfied by a quantized spacetime $N$ is the hyperbolic equation  
\begin{equation}\lae{3.1}
H_0{\tilde u}-H_1{\tilde u}=0
\end{equation}
in a quantum spacetime
\begin{equation}\lae{3.2}
Q=(0,\un)\times \so
\end{equation}
where $\so$ is a Cauchy hypersurface of $N$ with induced metric $g_{ab}$. $H_0$ is the temporal Hamiltonian, a temporal self-adjoint operator with countably many eigenfunctions $w_i$ and corresponding eigenvalues $\lam_i>0$ and $H_1$ the spatial Hamiltonian, which, in our case, is also a self-adjoint operator but with a continuous spectrum.

In order to solve the hyperbolic equation we have to solve the eigenvalue equation 
\begin{equation}\lae{3.3}
H_1v_i=\lam_i v_i,
\end{equation}
for every $i\in\N$, where
\begin{equation}\lae{3.4}
H_1v=\frac {16(n-1)}n \{-(n-1)\D v-\frac{n-2}2 Rv\}
\end{equation}
and the Laplacian and the scalar curvature refer to the induced metric of the Cauchy hypersurface in the exterior region of the black hole
\begin{equation}
\so=\{t=\const\}.
\end{equation}
Let us recall that the induced metric is independent of $t$ and the scalar curvature $R$ is constant
\begin{equation}\lae{3.6}
R=2\Lam,
\end{equation}
\cf \frl{2.1}. The solutions $\tilde u$ of \re{3.1} could then be expressed as products of respective eigenfunctions or eigendistributions
\begin{equation}
\tilde u_i=w_iv_i\qq\A\,i\in \N.
\end{equation}
The multiplicities of the temporal eigenfunctions $\lam_i$ are all equal to one in contrast to the multiplicities of the spatial eigenvalues as will become evident shortly. Hence we should label the spatial eigenfunctions in the form
\begin{equation}
v_{ij},\qq 1\le j\le m_i,
\end{equation}
such that
\begin{equation}\lae{3.9}
H_1v_{ij}=\lam_i v_{ij}, \qq\A\,1\le j\le m_i.
\end{equation}
The eigenfunctions $v$ in \frt{2.9} with eigenvalue $\mu$ are, obviously, also eigenfunctions of the operator $H_1$, since $R$ is constant. Hence, we conclude
\begin{equation}\lae{3.10}
\begin{aligned}
H_1v&= \frac {16(n-1)}n\{(n-1)\mu +(n-2)\abs\Lam\}v\\
&=\frac {16(n-1)}n (n-1)\{\mu +\frac{n-2}{n-1}\abs\Lam\}v
\end{aligned}
\end{equation}in view of \re{3.4} and \re{3.6}.

In order to solve equation \re{3.3} for all $i\in\N$ we have to prove that there exists an admissible $\mu\in \R[]$ such that 
\begin{equation}
\lam_0\ge \frac {16(n-1)}n (n-1)\{\mu +\frac{n-2}{n-1}\abs\Lam\},
\end{equation}
where $\mu$ is admissible provided
\begin{equation}
\mu>-\frac{n-1}{2n}\abs\Lam,
\end{equation}
\cf \fre{2.41}. Thus, we must verify that
\begin{equation}
\lam_0>\frac {16(n-1)}n (n-1)\{-\frac{n-1}{2n} +\frac{n-2}{n-1}\}\abs\Lam.
\end{equation}
On the other hand, we know
\begin{equation}
\lam_i=\bar\lam_i \abs\Lam ^\frac{n-1}n,
\end{equation}
where $\bar\lam_i$ is the eigenvalue corresponding to $\abs\Lam=1$, \cf \cite[Lemma 9.4.8, p.~240]{cg:qgravity-book2}, from which we deduce 
\begin{equation}
\bar\lam_0>\frac {16(n-1)}n (n-1)\{-\frac{n-1}{2n} +\frac{n-2}{n-1}\}\abs\Lam^\frac1n.
\end{equation}
Let $\Lam_0<0$ be defined by
\begin{equation}\lae{3.16}
\bar\lam_0=\frac {16(n-1)}n (n-1)\{-\frac{n-1}{2n} +\frac{n-2}{n-1}\}\abs{\Lam_0}^\frac1n,
\end{equation}
then we can prove:
\bt\lat{3.1}
Let $1\le m_i\in\N$ be a given sequence of multiplicities, then the eigenvalue problems \re{3.9} have solutions of the form 
\begin{equation}
v_{ij}=r^{-\frac{n-1}2} u_{ij}\f_{ij},
\end{equation}
where, for each $i\in\N$, $\f_{ij}=\f_{ij}(x)$ are mutually orthogonal eigenfunctions or linearly independent eigendistributions of equation \re{2.17} and $u_{ij}$ are linearly independent tempered eigendistributions of equation \re{2.35} with initial values \re{2.36}, provided the cosmological constant $\Lam$ satisfies
\begin{equation}\lae{3.18} 
\abs{\Lam_0}>\abs\Lam>0.
\end{equation}
\et
\bp
In view of the arguments preceding the theorem we have already proved that for each $i\in\N$ the functions $v_{ij}$ are eigenfunctions with eigenvalues $\lam_i$. It remains to prove that they are linearly independent but this is guaranteed by the choice of the $\f_{ij}$. Indeed, differentiating $v_{ij}$ with respect to $\tau$ and evaluating at $\tau=0$ yields
\begin{equation}
v_{ij}'(0,x)=r_0^{-\frac{n-1}2}\f_{ij}(x),
\end{equation}
where the right-hand side is linearly independent.

The linear independence of the $v_{ij}$ with mutually different $i$ is obvious, since the eigenvalues $\lam_i$ of $H_0$ have multiplicity one.
\ep

\section{The von Neumann entropy in the exterior region}\las{4}

In \cite[Chapter 9.5, p.~240]{cg:qgravity-book2} we have already defined the von Neumann entropy of the interior region of black holes, or more generally, of quantized globally hyperbolic spacetimes with a negative cosmological constant. Now we apply these definitions to the exterior region of a Schwarzschild-AdS black hole. For the sake of completeness let us repeat the necessary definitions and results without giving the proofs for which we refer to \cite[Chapter 9.5, p.~240]{cg:qgravity-book2}.

We first define the partition function by using the spatial Hamiltonian $H_1$ of the quantized black hole, which is now defined in the separable Hilbert space $\mc H$ generated by the eigendistributions
\begin{equation}
v_{ij},\qq \A\, i\in \N,\;1\le j\le m_i,
\end{equation}
which are sufficiently smooth functions satisfying the eigenvalue equations
\begin{equation}
H_1 v_{ij}=\lam_i  v_{ij}
\end{equation}
We also stipulate at the moment that the multiplicities $m_i$ are uniformly bounded, though later we are allowed to drop the requirement
\begin{equation}
m_i\le l,
\end{equation}
where $1\le l\in\N$ is arbitrary but fixed, \cf \rr{4.5}.

In order to explain how the eigendistributions can generate a Hilbert space let us relabel the eigenfunctions and the eigenvalues of $H_1$ by $(\tilde v_i,\tlam_i)$ such that
\begin{equation}
H_1 \tilde v_i=\tlam_i \tilde v_i,
\end{equation}
i.e., the multiplicities of the eigenvalues are now included in the labelling and the ordering is no longer strict
\begin{equation}
\tlam_0\le \tlam_1\le \tlam_2\le \cdots.
\end{equation}
To define the Hilbert space $\mc H$ we simply declare that the eigendistributions are mutually orthogonal unit eigenvectors, hence defining a scalar product in the complex vector space $\mc H'$ spanned by these eigenvectors by stipulating that the first entry of the scalar product should be antilinear. We define the Hilbert space $\mc H$ to be its completion.
\bl\lal{4.1}
The linear operator $H_1$ with domain $\mc H'$ is essentially self-adjoint in $\mc H$. Let $\bar H_1$ be its closure, then the only eigenvectors of $\bar H_1$ are those of $H_1$.
\el
\br
In the following we shall write $H_1$ instead of $\bar H_1$. 
\er
\bl
For any $\bet>0$ the operator
\begin{equation}
e^{-\bet H_1}
\end{equation}
is of trace class in $\mc H$. Let
\begin{equation}
\msc F\equiv\msc F_+(\mc H)
\end{equation}
be the symmetric Fock space generated by $\mc H$ and let
\begin{equation}
H=d\C(H_1)
\end{equation}
be the canonical extension of $H_1$ to $\msc F$. Then
\begin{equation}
e^{-\bet H}
\end{equation}
is also of trace class in $\msc F$
\begin{equation}\lae{4.12}
\tr(e^{-\bet H})=\prod_{i=0}^\un(1-e^{-\bet \tlam_i})^{-1}<\un.
\end{equation}
\el
We then define the partition function $Z$ by
\begin{equation}
Z=\tr(e^{-\bet H})=\prod_{i=0}^\un (1-e^{-\bet\tlam_i})^{-1}
\end{equation}
and the density operator $\rho$ in $\msc F$ by
\begin{equation}
\rho=Z^{-1}e^{-\bet H}
\end{equation}
such that
\begin{equation}
\tr \rho=1.
\end{equation}

The von Neumann entropy $S$ is then defined by
\begin{equation}\lae{4.16}  
\begin{aligned}
S&=-\tr(\rho\log \rho)\\
&=\log Z+\bet Z^{-1}\tr (He^{-\bet H})\\
&=\log Z-\bet\pde{\log Z}\bet\\
&\equiv \log Z +\bet E,
\end{aligned}
\end{equation}
where $E$ is the average energy
\begin{equation}
E=\tr (H\rho).
\end{equation}
$E$ can be expressed in the form
\begin{equation}\lae{4.18}
E=\sum_{i=0}^\un \frac{\tlam_i}{e^{\bet\tlam_i}-1}.
\end{equation}
Here, we also set the Boltzmann constant
\begin{equation}
K_B=1.
\end{equation}
The parameter $\bet$ is supposed to be the inverse of the absolute temperature $T$
\begin{equation}
\bet=T^{-1}.
\end{equation}
\bl\lal{4.4} 
The average energy $E$ and the entropy $S$ increase if the multiplicities $m_i$ of the eigenvalues increase. If we pick a fixed index $k$ and let the corresponding multiplicity $m_k$ converge to infinity, then the energy $E$ and the entropy $S$ tend to infinity
\begin{equation}
\lim_{m_k\ra\un}E=\un\q\wed\q \lim_{m_k\ra\un}S=\un.
\end{equation}
\el
\bp
In view of \re{4.18} we infer
\begin{equation}
E=\sum_{i=0}^\un m_i \frac{\lam_i}{e^{\bet\lam_i}-1}
\end{equation}
and
\begin{equation}
S\ge \bet E,
\end{equation}
due to equation \re{4.16} and the fact $Z> 1$, \cf \re{4.12}, where we recall that the eigenvalues are strictly positive. The claims of the lemma are now easily deduced.
\ep
We therefore are in a dilemma to define a physically reasonable average energy and entropy because theoretically we can increase the multiplicities $m_i$ arbitrarily, as we proved in \rt{3.1}, because for each $i$ there are countably many mutually orthogonal eigenfunctions $\f_{ij}$ satisfying equation \fre{2.17} which can be used to define a solution $u_{ij}$ of the eigenvalue equation \fre{2.35} with eigenvalue $\mu=\lam_i$ without changing the value of that eigenvalue. The eigenvalues of $\f_{ij}$ only determine the coefficients of the lower order terms of the operator $A_0$ which has a continuous spectrum which is independent of these eigenvalues. Hence, we cannot develop a meaningful quantum statistics for the exterior region by looking at it as an isolated system. Instead we also have to take the interior region of the black hole into account.

In \cite[Chapter 7.1]{cg:qgravity-book2} we quantized the interior region of a Schwarzschild-AdS black hole, where $x^0=r$, $0<r<r_0$, is the time function and the Cauchy hypersurfaces are the slices
\begin{equation}
S_r=\{x^0=r\}.
\end{equation}
The multiplicities of the eigenvalues of the corresponding spatial Hamiltonian $H_1$, which we denoted by $n(\lam_i)$, could not be chosen arbitrarily large; instead they were  defined by maximizing their values. The $n(\lam_i)$ also depended monotonically increasingly on $r$, and we proved that the $S_r$, $0<r<r_0$, equipped with the induced metric converged to a smooth compact Riemannian manifold $S_{r_0}$, if $r$ tended to $r_0$, which we considered to be the horizon. If $\tilde \ka\not=-1$, then the results were valid for any $\Lam<0$; only in case $\tilde\ka =-1$ we had to assume 
\begin{equation}\lae{4.23.1} 
\abs \Lam\ge \abs{\Lam_0}>0
\end{equation}
for some specific value $\abs{\Lam_0}$ in order to find a spatial eigenfunction for the smallest eigenvalue $\lam_0$, \cf \cite[equ. (7.2.30), p.~179]{cg:qgravity-book2}. Furthermore, we proved that, for $\bet>0$,
\begin{equation}
e^{-\bet H_1}
\end{equation}
is of trace class, \cf \cite[pp.~231--236]{cg:qgravity-book2}, such that quantum statistics could be applied to this configuration.
\br\lar{4.5}
Note that the $n(\lam_i)$ are not uniformly bounded. Indeed from the equations \cite[equs. (9.4.60), (9.4.72), pp.~234--235]{cg:qgravity-book2} we conclude 
\begin{equation}
\lim_{i\ra \un}n(\lam_i)=\un.
\end{equation}
\er

Using this result we combine the quantization of the exterior region with the quantization of the interior region by defining 
\begin{equation}\lae{4.28}
m_i=n(\lam_i),\qq\A\, i\in\N,
\end{equation}
such that the quantum statistical results are identical for both regions, though the spatial eigenfunctions are of course different. Let us summarize this result in a theorem. 
\bt\lat{4.6}
Let $N=N^{n+1}$, $n\ge 3$, be a Schwarzschild-AdS black hole with metric defined in \re{2.1} and \fre{2.2} and assume that $\tilde\ka\not=-1$. The negative cosmological constant $\Lam$ should satisfy
the estimate \fre{3.18}. Then, choosing an arbitrary Cauchy hypersurface $\so =S_t$ in the exterior region, this region can be quantized by canonical quantum gravity such that the hyperbolic equation \re{3.1} in the quantum spacetime \re{3.2} can be solved by a sequence of products of temporal eigenfunctions and spatial tempered eigendistributions which generate a Hilbert space $\mc H$ and can be looked at as being mutually orthogonal unit vectors.

 A similar result is also valid in the interior region of the black hole. By choosing the multiplicities $m_i$ of the spatial eigenvalues $\lam_i$ in the exterior region to be identical to the multiplicities $n(\lam_i)$ of the spatial eigenvalues in the interior region, which seems to be the only logical possibility, we conclude that the quantum statistical results in both regions are identical, i.e., the respective partition functions, average energies and entropies are identical.  
 
 Furthermore, since the spatial eigendistributions in both regions are an orthonormal basis in their respective Hilbert spaces, the relation \re{4.28} can be used to define a unitary map between these Hilbert spaces such that the respective spatial Hamiltonians are unitarily equivalent because the corresponding eigenvalues are identical. Hence, the black hole information paradox does not exist on a quantum level.
\et
\br
If  we want to extend the above theorem to the case $\tilde \ka=-1$, then $\Lam$ would have to satisfy both the condition \fre{3.18} as well as \re{4.23.1}, where in the latter case $\Lam_0$ is different, \cf [4, equ. (7.2.29), p. 178]. Let us call this $\Lam_0$ $\tilde\Lam_0$. Then $\Lam$ would have to satisfy
\begin{equation}
\abs{\Lam_0}>\abs{\Lam}\ge \abs{\tilde\Lam_0}. 
\end{equation}
One would have to check if 
\begin{equation}
\abs{\Lam_0}>\abs{\tilde\Lam_0}.
\end{equation}
If this is the case, then for the $\abs\Lam$ belonging to this interval unitary equivalence would be applicable.
\er
\section{Schwarzschild-dS black holes}

If the cosmological constant $\Lam$ is positive then the Schwarzschild-dS spacetime has a singularity in $r=0$ only if $\tilde\ka=1$. Then the metric in the exterior region of the black hole can be expressed as in \fre{2.1}. Since we want to quantize the black hole by similar arguments as in case $\Lam<0$, especially with regard to the temporal eigenfunctions, we also have to assume $n=3$.
In this case the quantum development of a Cauchy hypersurface $\so$ is governed by the equation
\begin{equation}\lae{4.2.57.4.1}
\begin{aligned}
&\frac14\frac n{16(n-1)}t^{-(m+k)}\frac\pa {\pa t}(t^{(m+k)}\dot u)\\
&\q -\frac14t^{-2}\D_Mu+\frac78 t^{-2}\D_{\R[k]}u-(n-1)t^{2-\frac4n}\tilde\D_\s u\\
&\q+\frac14t^{2-\frac4n}R_\s u -\frac12t^2\Lam u=0,
\end{aligned}
\end{equation}
\cf \cite[equ.~(4.2.57), p.~115]{cg:qgravity-book2}, which is structurally identical to equation \fre{6.1.1} because $\Lam$ is positive. Solving this equation by temporal and spatial eigenfunctions can be achieved by identical arguments as in case $\Lam<0$ having in mind that $n=3$ and $\tilde\ka=1$. 
 
\section{Conclusion}\las{6}
We quantized the exterior region of a Schwarzschild-AdS black hole and solved the resulting hyperbolic equation
\begin{equation}
H_0\tilde u_{ij}-H_1 \tilde u_{ij}=0
\end{equation}
in the quantum spacetime
\begin{equation}
Q=(0,\un)\times \so
\end{equation}
by a sequence of eigenfunctions, or eigendistributions
\begin{equation}
\tilde u_{ij}=w_i v_{ij},\qq 1\le j\le m_i,
\end{equation}
where $w_i$ are the eigenfunctions of the temporal Hamiltonian $H_0$ and $v_{ij}$ are the eigendistributions of the spatial Hamiltonian $H_1$. The corresponding eigenvalues are $\lam_i$, where the multiplicities of the eigenvalues are one for $H_0$ and $m_i$ for $H_1$. For fixed $i$ the multiplicity $m_i$ is theoretically unbounded. There are no physical or mathematical obstructions which would enforce a bound from above. Since the von Neumann entropy and the average energy tend to infinity if $m_i$ converges to infinity, \cf \frl{4.4}, the only logical and physical consequence is to define $m_i$ to be equal to the corresponding multiplicity $n(\lam_i)$ which we obtained by maximizing the multiplicities in the interior case, where we quantized the interior region of a Schwarzschild-AdS black hole, \cf \cite[Chapter 7]{cg:qgravity-book2}. The $v_{ij}$ can be looked at as mutually orthogonal eigenvectors generating a complex Hilbert space as in the interior case. Hence, setting $m_i=n(\lam_i)$ implicitly defines a unitary map between the respective Hilbert spaces and the respective Hamiltonians are unitarily equivalent, since the eigenvalues are also identical. Thus, the partition function, the von Neumann entropy and the average energy in the respective regions are all identical and there does not exist an information paradox, \cf the papers \cite{hawking:information,hawking:information2,page:information,page:information2,almheiri:entropy}. 

\bt
The spatial eigenfunctions $v_{ij}=v_{ij}(\tau,x)$ can be looked at as being gravitational waves emanating from the event horizon and vanishing exponentially fast at infinity satisfying   
\begin{equation}
v_{ij}(0,x)=0
\end{equation}
and
\begin{equation}\lae{6.10}
\lim_{\tau\ra \un} \abs{v_{ij}(\tau,\cdot)}_{m,M_0}=0\qq\A\,m\in \N,
\end{equation}
where we use the norm in $C^m(M_0)$. Furthermore, $v'_{ij}$ also vanishes exponentially fast at infinity such that
\begin{equation}\lae{6.11}
\sup_{x\in M_0}\int_0^\un(\abs{v_{ij}(\tau,x)}^2+\abs{v'_{ij}(\tau,x)}^2)\,d\tau<\un.
\end{equation}
\et
\bp
Let us recall the definition of $\tau$ in \fre{2.11}, the estimate \fre{2.44},
\begin{equation}
r^{-1}(\tau)\le c_2 e^{-c_1\tau},
\end{equation}
and the definitions of $v_{ij}$ and $u_{ij}$ in \frt{3.1},
\begin{equation}
v_{ij}(\tau,x)=r^{-\frac{n-1}2}(\tau) u_{ij}(\tau)\f_{ij}(x),
\end{equation}
where $\f_{ij}$ are the smooth eigenfunctions of the Laplacian of the compact space form $M_0$ and the eigendistributions $u_{ij}$ have the initial values  
\begin{equation}
u_{ij}(0)=0
\end{equation}
and
\begin{equation}
u_{ij}'(0)=1
\end{equation}
and satisfy the estimate
\begin{equation}
\abs{u_{ij}(\tau)}+\abs{u_{ij}'(\tau)}\le c\qq\A\,\tau\in [0,\un),
\end{equation}
\cf \frt{2.6} and \frc{2.8}.
The proof of the claims, except the behavior of $v'_{ij}$ and inequality \re{6.11}, is now straightforward because $r(0)=r_0$, the radius of the event horizon.

\cvm
\nd 
In order to prove the asymptotic behavior of $v'_{ij}$ and inequality \re{6.11} we differentiate $v_{ij}$ with respect to $\tau$
\begin{equation}
\begin{aligned}
v'_{ij}&=-\frac{n-1}2 r^{-\frac{n-1}2} r^{-1}r'u_{ij}\f_{ij}+r^{-\frac{n-1}2} u'_{ij}\f_{ij}\\
&=-\frac{n-1}2 r^{-\frac{n-1}2} r^{-1}h^\frac12 u_{ij}\f_{ij}+r^{-\frac{n-1}2} u'_{ij}\f_{ij},
\end{aligned}
\end{equation}
where we used the relation \fre{2.12}. Moreover, from the definition of $h$ in \re{2.2} we deduce 
\begin{equation}
r^{-1} h^\frac12\le c\qq\A\, r\ge r_0
\end{equation}
with a uniform constant $c$. Since $u_{ij}, \,u'_{ij}$ and $\f_{ij}$ are also uniformly bounded the exponential decay of $v'_{ij}$ and inequality \re{6.11} can be inferred from \re{2.44}. 
\ep

%\backmatter
%\includepdf[pages=-]{/Users/claus/Documents/Scanned-Documents/}
\bibliographystyle{hamsplain}
%\bibliography{mrabbrev,publications}

\begin{thebibliography}{10}

\bibitem{almheiri:entropy}
Ahmed Almheiri, Thomas Hartman, Juan Maldacena, Edgar Shaghoulian, and
  Amirhossein Tajdini, \emph{{The entropy of Hawking radiation}}, Rev. Mod.
  Phys. \textbf{93} (2021), 035002,
  {\href{https://doi.org/10.1103/RevModPhys.93.035002}{doi:10.1103/RevModPhys.93.035002}}.

\bibitem{cg:cp}
Claus Gerhardt, \emph{Curvature {P}roblems}, Series in Geometry and Topology,
  vol.~39, International Press, Somerville, MA, 2006.

\bibitem{cg:qbh2}
Claus Gerhardt, \emph{{The quantization of a Kerr-AdS black hole}}, Advances in
  Mathematical Physics \textbf{vol. 2018} (2018), Article ID 4328312, 10 pages,
  {\href{http://arXiv.org/abs/1708.04611}{arXiv:1708.04611}},
  {\href{http://dx.doi.org/10.1155/2018/4328312}{doi:10.1155/2018/4328312}}.

\bibitem{cg:qgravity-book2}
Claus Gerhardt, \emph{{The Quantization of Gravity}}, 2nd ed., Fundamental Theories of
  Physics, vol. 194, Springer, Cham, November 2024,
  {\href{https://doi.org/10.1007/978-3-031-67922-3}{doi:10.1007/978-3-031-67922-3}}.

\bibitem{cg:qgravity5}
Claus Gerhardt, \emph{{Extending the solutions and the equations of quantum gravity
  past the big bang singularity}}, Symmetry \textbf{17} (2025), no.~2, 262,
  {\url{https://doi.org/10.3390/sym17020262}}.

\bibitem{hawking:information}
S.~W. Hawking, \emph{{Breakdown of predictability in gravitational collapse}},
  Phys. Rev. D \textbf{14} (1976), 2460--2473,
  {\href{https://doi.org/10.1103/PhysRevD.14.2460}{doi:10.1103/PhysRevD.14.2460}}.

\bibitem{hawking:information2}
S.~W. Hawking, \emph{Information loss in black holes}, Phys. Rev. D \textbf{72}
  (2005), 084013,
  {\href{https://doi.org/10.1103/PhysRevD.72.084013}{doi:10.1103/PhysRevD.72.084013}}.

\bibitem{simon-spectrum}
Alexander Kiselev, Yoram Last, and Barry Simon, \emph{{Modified Pr{\"u}fer and
  EFGP Transforms and the Spectral Analysis of One-Dimensional Schr{\"o}dinger
  Operators}}, Communications in Mathematical Physics \textbf{194} (1998),
  no.~1, 1--45,
  {\href{https://doi.org/10.1007/s002200050346}{doi:10.1007/s002200050346}}.

\bibitem{maurin:book}
Krzysztof Maurin, \emph{Methods of {H}ilbert spaces}, Translated from the
  Polish by Andrzej Alexiewicz and Waclaw Zawadowski. Monografie Matematyczne,
  Tom 45, Pa\'nstwowe Wydawnictwo Naukowe, Warsaw, 1967.

\bibitem{page:information}
Don~N. Page, \emph{Information in black hole radiation}, Phys. Rev. Lett.
  \textbf{71} (1993), 3743--3746,
  {\href{https://doi.org/10.1103/PhysRevLett.71.3743}{doi:10.1103/PhysRevLett.71.3743}}.

\bibitem{page:information2}
Don~N Page, \emph{{Time dependence of Hawking radiation entropy}}, Journal of
  Cosmology and Astroparticle Physics \textbf{2013} (2013), no.~09, 028--028,
  {\href{https://doi.org/10.1088/1475-7516/2013/09/028}{doi:10.1088/1475-7516/2013/09/028}}.

\end{thebibliography}
%\providecommand{\bysame}{\leavevmode\hbox to3em{\hrulefill}\thinspace}
\providecommand{\href}[2]{#2}

%\listoffigures

%\cleardoublepage

%\thispagestyle{empty}
%\closegraphsfile
\end{document}